\documentclass[11pt,a4paper,doublespace,onecolumn]{aastex}
\usepackage{graphicx}
\usepackage{graphicx}
\usepackage{lineno}

\begin{document}

\title{Evolution of induced earthquakes from dimensionless scaling}

\author{Maurice H.P.M. van Putten}
\affil{Astronomy and Astrophyscis, Sejong University, 98 Gunja-Dong, Gwangjin-gu, 143-747 Seoul, South Korea}
\author{Anton F.P. van Putten}
\affil{AnMar Research Laboratories B.V., 5645 KK Eindhoven, The Netherlands}
\author{Michel J.A.M. van Putten}
\affil{Clinical neurophysiology, Medisch Spectrum Twente and University of Twente, PO Box 217, 7500 AE Enschede, The Netherlands}

\begin{abstract}
Event counts are a powerful diagnostic for earthquake analysis. 
We report on a bimodal distribution of earthquake event counts in the U.S. since 2013. The new peak is about a magnitude M1 with a distribution very similar to that of induced earthquakes in Groningen, The Netherlands. The event count of the latter shows exponential growth since 2001 with a doubling time of 6.24 years  at a relatively constant rate of land subsidence. We attribute this to internal shear stresses due to increased curvature. We demonstrate exponential growth as a function of dimensionless curvature in a tabletop crack formation experiment. An outcome of our model is a new method of parameter-free statistical 
forecasting of induced events, that circumvents the need for a magnitude cut-off in the Gutenberg-Richter relationship.
\end{abstract}

\maketitle

\section{Introduction}

Increases in earthquake event rates are a modern phenomenon that may result from
exploitation of natural energy resources or injection of waste \citep{ral76,nic90,fro12,eva12}. These increases appear as 
discontinuities in event count rates in excess over historical trends in seismic activity. In the period 2001-2012, 
the central and eastern United States experienced an excess of over 300 events with magnitude greater than 3.0 above what would be expected based on natural seismic activity alone \citep{ell13}. Fig. 1 shows what appears to be a similar increase in 
event rates in The Netherlands, generally attributed to the exploitation of the Groningen gas field. 

Although induced earthquakes tend to be moderate in magnitude on average, they can, by their increasing frequency,
act as physical stressors producing gradual degradation of the structural integrity of homes and buildings.
These events can become endemic, when acting as a psychological stressor to a local population. Due to the earthquakes 
in Northeast Groningen, half of the inhabitants are now living with a feeling of anxiety or fear and wish to move out \citep{kam14}. Most likely, the psychological impact caused by earthquakes is strengthened by their stochastic nature \citep{fon09,gri04}. With a population of about half a million inhabitants, the ubiquity of earthquakes in 
Groningen is currently a topic of considerable concern for the development of ameliorating policies (e.g. \cite{vol14}). However, ensuring safety of the population and preventing further damage to buildings by reducing earthquake events is challenging in view of the uncertainties in statistical forecasting \citep{nic90,bak05,nam13}.

The Groningen gas field is monitored for reservoir pressure, subsidence and yearly production of natural gas covering an 
area of about 900 km$^2$ \citep{nam13}. Since 2001, the present trend in earthquake event counts is essentially exponential,
shown by the red line in the cumulative event count in Fig. 1. (In case of exponential growth, cumulate event counts and event count rates both appear linearly on a logarithmic scale.) This type of `clock-work' precision is robust in persisting over a broad range of the cut-off in magnitude, here expressed on the scale of Richter, showing a doubling time of 6.24 years, slightly above a previous estimate of 5 years \citep{dos13}). The Groningen field, therefore, may be showing us a principle at work with possible implications for the interpretation of earthquake surveys elsewhere.

The Groningen gas field has been gradually developed over the last few decades. Fig. 2 shows the production levels of the
Nederlandse Aardolie Maatschappij (NAM) in GNm$^3$ per year from 1990-2011. 
As a result, reservoir pressures have steadily decreased to below 100 bar at present from about 150 bar in 2000 and starting at 350 bar at discovery in 1959. Over the last decade, the exploitation to the present level of about 50 billion Nm$^3$ per year evolved with variations up to about 20\% around a linear fit. The standard deviation of the relative variations in NAM production about this linear trend is 9.4\%. There is no discernible correlation between the small fluctuations in the logarithm of the event counts to production. 

Quite generally, earthquakes can be induced by reduction in well pressure or raising shear stresses \citep{hub59}. They may occur at some distance away from industrial activity and with a considerable delay in time. Reviewed in \citep{ell13}, Paradox Valley, Colorado, shows events at over ten kilometers away from injection wells with delays of well over a decade. The Salton Sea Geothermal Field shows, after some delay following the starting date of exploitation, a remarkable correlation of induced seismicity with net extracted volume $\dot{V}$ of a few times $10^6$ m$^3$ per month at a depth of about 1-2.5 km, defined as extracted minus injected fluid volume \citep{bro13}. These observations suggest that induced seismic activity is associated with large area deformations measured in subsidence. 

The observed subsidence $\dot{h}$ of 0.5 to 2 cm per year in the Groningen area is a delayed response to 
the volumetric reduction in a gradual compactification in the reservoir \citep{nam13}. The associated volume reduction 
$\dot{V}\simeq 10^6$ m$^{-3}$ per month is, within one order of magnitude, quantitatively similar to that in the Salton 
Sea Geothermal Field. Soil subsidence reflects a perturbation of internal stresses on large scales, particularly shear 
stresses associated with curvature induced strain, that may reach critical values for material failure or slip, e.g., as 
described by the Mohr-Coulomb law \citep{moh76,hub59}. In the Groningen area, subsidence is presently taking 
the shape of a shallow bowl with a radius $r\simeq$ 15 km with a current depth of about 30 cm (\citep{nam13},
schematically illustrated in Fig. 3 below). Relative to the radius of curvature, $R$, it presents a dimensionless curvature 
$\rho=r/R\simeq 4\times 10^{-5}$.

\begin{figure}[h]
\centerline{\includegraphics[scale=0.475]{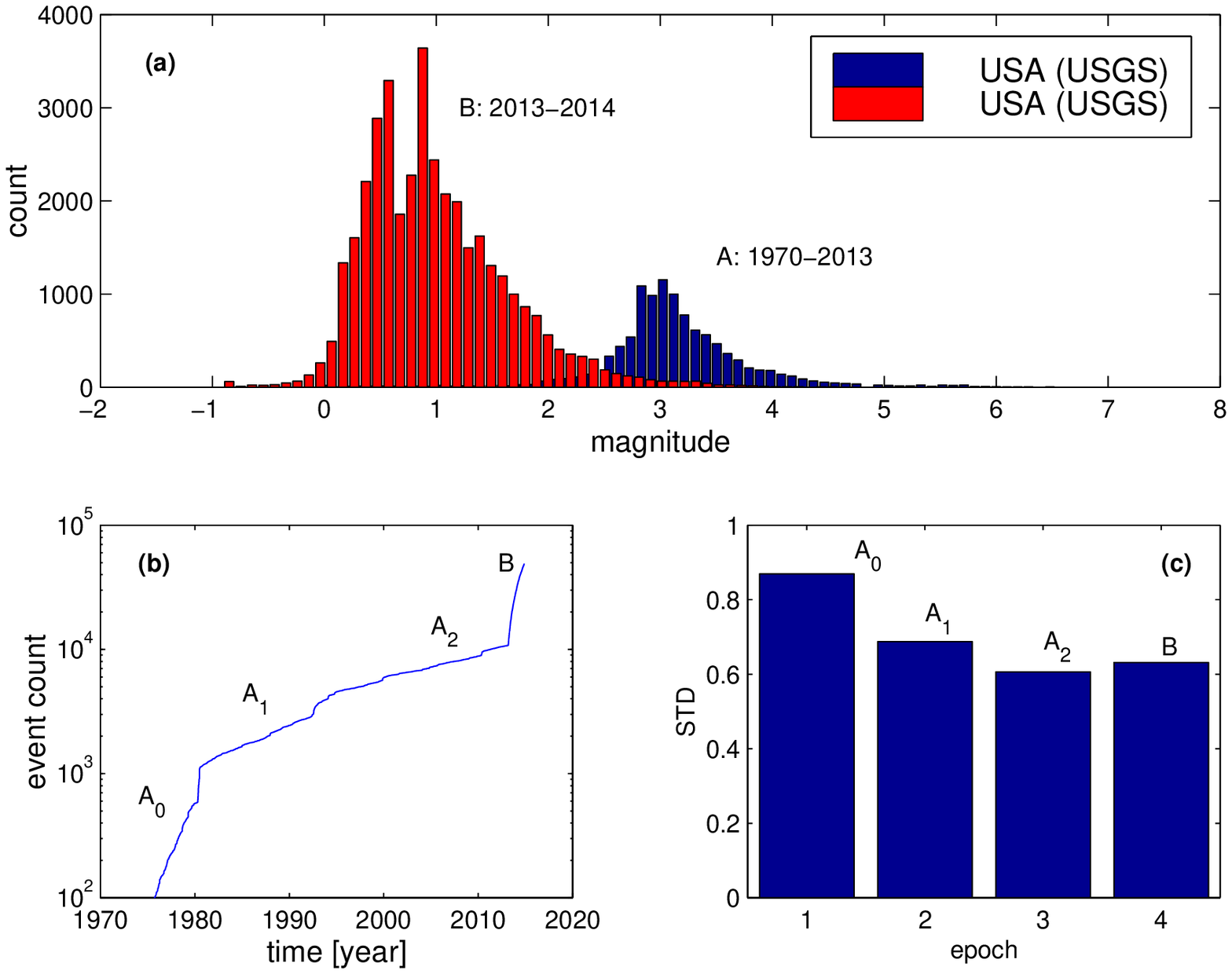}\includegraphics[scale=0.475]{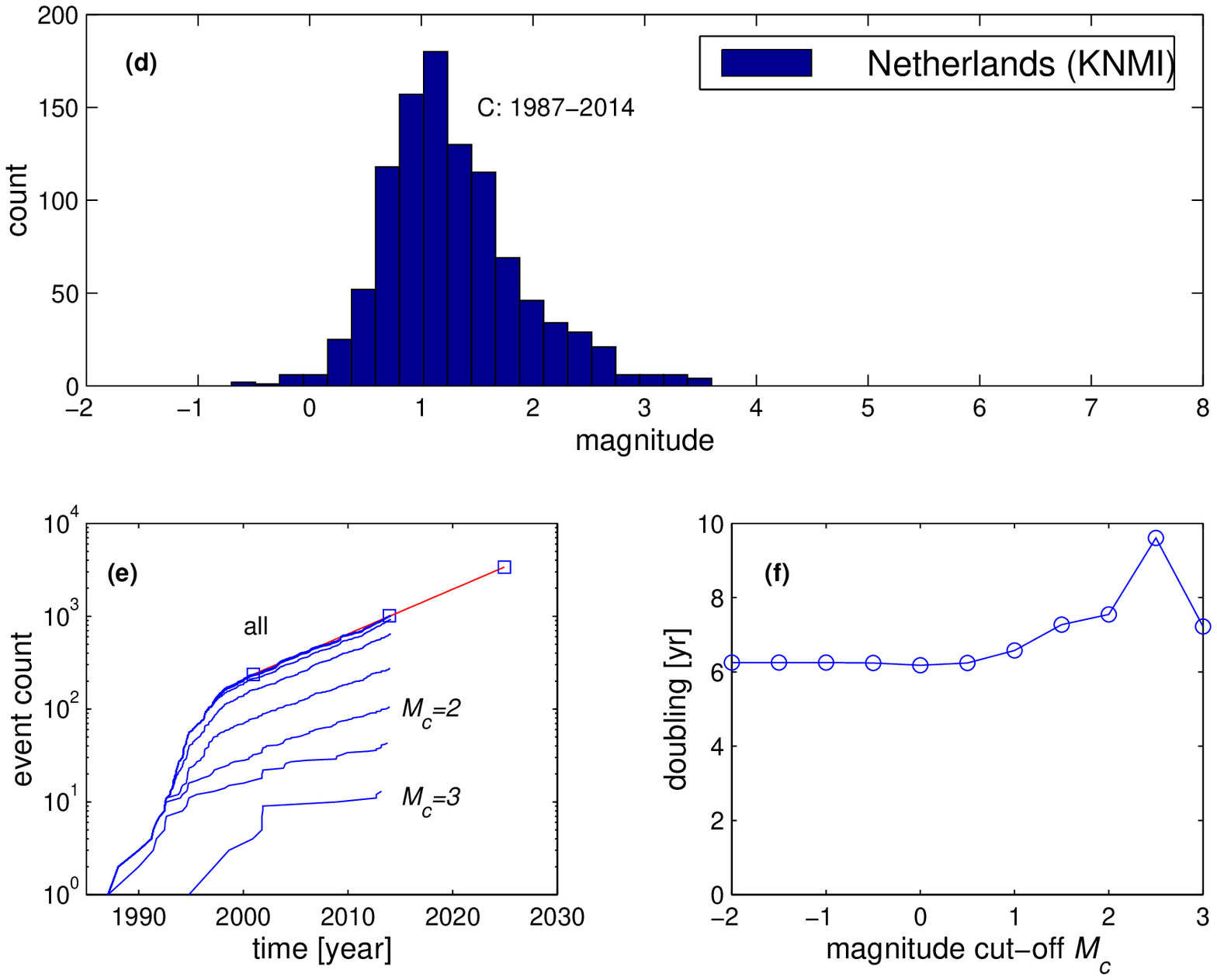}}
\caption{(a) Earthquake event counts in the United States with depth less than 4 km (a). (d) Event
counts of induced events in The Netherlands. The distribution in the US  is now bimodal with $B$ entirely due 2013-2014. 
Events in $B$ in the US and $C$ in the Netherlands have similar STDs ($\sigma_B=0.6312$, $\sigma_C=0.6131$),
pointing to a common origin as induced events. (b,c) During 1980-2013, the logarithm of event counts in the US shows various 
transitions to approximately linear trends in $A_1$ and $A_2$. (e,f) The same such trend is particularly pronounced in $C$,
which accounts for a relatively short doubling time of 6.24 years, pointing to one event per day in 2025.}
\label{fig:GC1}
\end{figure}
\begin{figure}[h]
\centerline{\includegraphics[scale=0.5]{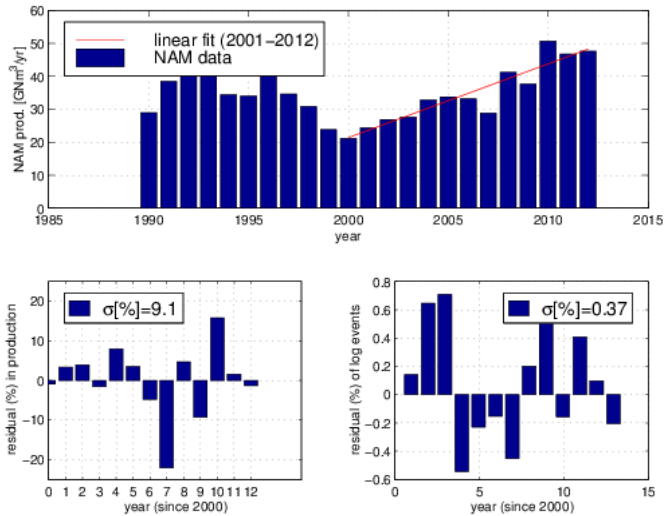}}
\caption{({\em Top}.) Overview of NAM production and deviations from a linear least squares fit in the period of 2000-2012. ({\em Bottom.}) Deviations about the linear fit are up to 20\% with a standard deviation of 9.4\%. Shown are also the deviations from linear trend in the logarithm of the event count, plotted in time. Fluctuations in the event rate reveal no correlation to (man-made) variations in NAM production.} 
\label{fig:GC2}
\end{figure}

Here, we present a physical mechanism for exponential growth in event counts from subsidence, resulting from curvature induced strain. The model gives a derivation of the associated Gutenberg-Richter relationship for statistical forecasting of the expected and maximum magnitude of events. We will apply our model to the Groningen data for a forecast up to 2025.

\section{Exponential growth from large scale deformations}

In the absence of discontinuities in cracks and slides, small perturbations in internal stress scale linearly with strain induced at 
a well. Strain then satisfies a linear elliptical equation, since propagation times of sound waves are negligible on the length scales of interest. The solution is determined by boundary conditions at the surface, where the tangential surface stress vanishes with pressure equilibration to ambient pressure. At the depth of the reservoir, e.g., $h\simeq$3 km in Groningen, the strain and pressure gradients are maximal at the well. In the presence of subsidence, the internal stresses are hereby a solution to a moving boundary problem. For relatively small variations in subsidence $\delta h/h<<1$, these stresses hereby scale linearly with subsidence $\delta h$ over the bulk of a region of volume $\sim h\times L^2$, where $L$ is the length scale of the reservoir.

\begin{figure}[h]
\centerline{\includegraphics[scale=0.35]{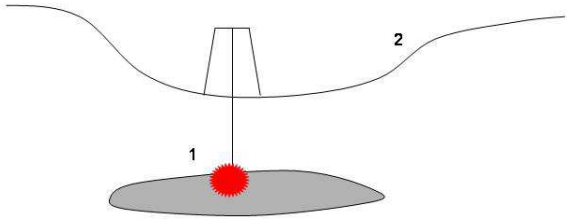}}
\caption{Schematic view of the geometry of the elliptic problem for the strain in the region around 
a well ({\em red}) in the exploitation of a reservoir ({\em grey}). The problem is a linear moving boundary problem 
for small perturbations, subject to the ambient pressure and vanishing shear stresses at the surface and 
vanishing perturbations at asymptotically large depth. In a dimensional analysis of induced earthquakes, the 
moving boundary is effectively parameterized by its dimensionless curvature. }
\label{fig:S0}
\end{figure}

For arbitrary perturbations, however, the problem is nonlinear since internal shear stresses are susceptible to sudden discontinuities when exceeding a critical value $\epsilon_c$, commonly described by the Mohr-Coulomb law of slippage. $\epsilon_c$ is  dimensionless, representing a critical ratio $\delta L/L \simeq \rho=\delta h/h$. A large scale deformation, $\delta L/L$ may be detected in subsidence $\delta h < 0$. As a moving boundary, $\rho$ effectively tracks the evolution of internal shear stresses (Fig. \ref{fig:S0}). Earthquakes are triggered when $\epsilon \ge \epsilon_c$, in the energetic release of stresses.

\begin{figure}[h]
\centerline{\includegraphics[scale=0.35]{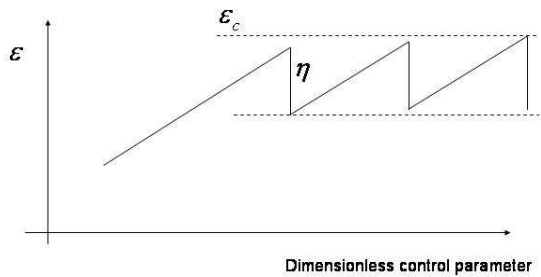}}
\caption{Schematic view of the local evolution of the dimensionless strain as a function of a dimensionless control
parameter such as relative subsidence $\delta h/h$ or dimensionless curvature $\rho$. Strain is limited due to
the formation of cracks or slides in accord with the Mohr-Coulomb law. A crack or slide defines a discontinuity in
momentum, giving rise to a release of energy followed by settling down to a sub-critical strain. This relaxation
is described by a jump $\eta$ in $\epsilon$. Shown is an average $\eta$, relevant to cracks or slides occurring 
subsequently at randomly different sites.}
\label{fig:T0}
\end{figure}

\begin{figure}[h]
\centerline{\includegraphics[scale=0.47]{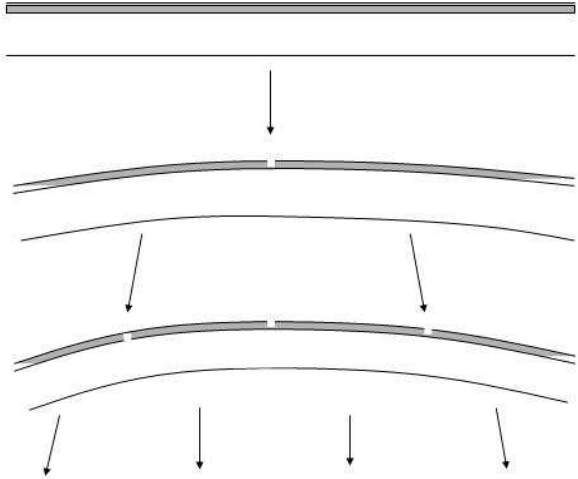}}
\caption{A schematic view of a tabletop experiment on crack formation in a metal foil glued to a flexible substrate,
subject to bending. The first crack induced by curvature forms at a critical strain $\epsilon=\epsilon_c$, which 
partitions the foil into two sections followed by relaxation to a stressed but stable state at a strain $\epsilon_c-\eta$. 
Further bending builds up stress back to $\epsilon_c$, causing the two sections to crack once more with subsequent
relaxation. The result is exponential growth in the number of cracks as a function of dimensionless curvature.}
\label{fig:T1}
\end{figure}

A sudden crack or slide partitions a region into two. It is followed by a relaxation $\eta$, $ 0 < \eta < \epsilon_c$, that determines the strain after the crack comes to a halt in a sub-critical but stressed state. 
If subsidence $\delta h < 0$ continues, strain increases again, possibly back up to 
$\epsilon_c$ from $\epsilon_c-\eta$. In approaching $\epsilon_c$ once more, regions in the new partition crack or slice into two once again. Effectively, the time-evolving strain $\epsilon=\epsilon(t)$ is restricted to the range (Fig. \ref{fig:T0})
\begin{eqnarray}
\epsilon_c-\eta < \epsilon < \epsilon_c.
\label{EQN_eps}
\end{eqnarray}
A persistent increase in large scale deformation introduces a continuously growing dimensionless curvature, $\rho$, as may be observed in subsidence and the association formation of a shallow bowl. The $\rho$ induced strain $\epsilon$ hereby extends over the macroscopic length scale $L$ and approximately uniformly so. The resulting evolution of partitions subject to (\ref{EQN_eps}), therefore, is one of exponential growth (Fig. \ref{fig:T1})
\begin{eqnarray}
N \propto e ^{\alpha \rho}
\label{EQN_N}
\end{eqnarray}
between event counts, where $\alpha$ is a material dependent dimensionless constant. 

Fig. 6 shows a tabletop demonstration of exponential growth in cracks. Here, shear stresses in excess of $\epsilon_c$ are 
made visible in cracks in aluminum foil glued to a flexible substrate as a function of the dimensionless curvature $L/R$, where
$L\simeq$ 20 cm is the length of the bar and $R$ is the radius of curvature. The experiment demonstrates exponentiation in
cracks driven by large scale deformations, here parametrized by curvature. In the field, large scale land deformations are
commonly parametrized by subsidence with associated curvature evident in bowl shape indentation centered around 
the well.

The above scaling argument for exponential growth in the number of cracks or slides with land deformation is based on the
dimensionless critical parameters $\epsilon_c$ and $\eta$ with no reference to an internal scale of length, area or volume.
Exponentiation of earthquake number is hereby independent of dimension, by which our one dimensional experiment is
meaningful also in three dimensions such as microfracturing experiments \citep{sch68}.

The observed subsidence rate in the Groningen field is remarkably constant in time \citep{nam13}. According to our mechanism, the number cracks or slides grows exponentially with subsidence. At a constant rate of the latter, the number of earthquakes hereby evolves exponentially in time, which accounts for the observed trend in the Groningen event count since 2001 (Fig. 1).

Our model describes a correlation between event counts and observed curvature, but does not address the problem
of time delay, between the activity at the well and the evolution of subsidence. Time delays are notoriously varied
in different locations, and their origin remains somewhat mysterious especially when delays are long, e.g., decades.
In case of the Groningen gas field, we were not able to constrain or determine the time delay, given a complete lack of  
correlation between gas exploitation and event counts.

\begin{figure}
\centerline{\includegraphics[scale=0.4]{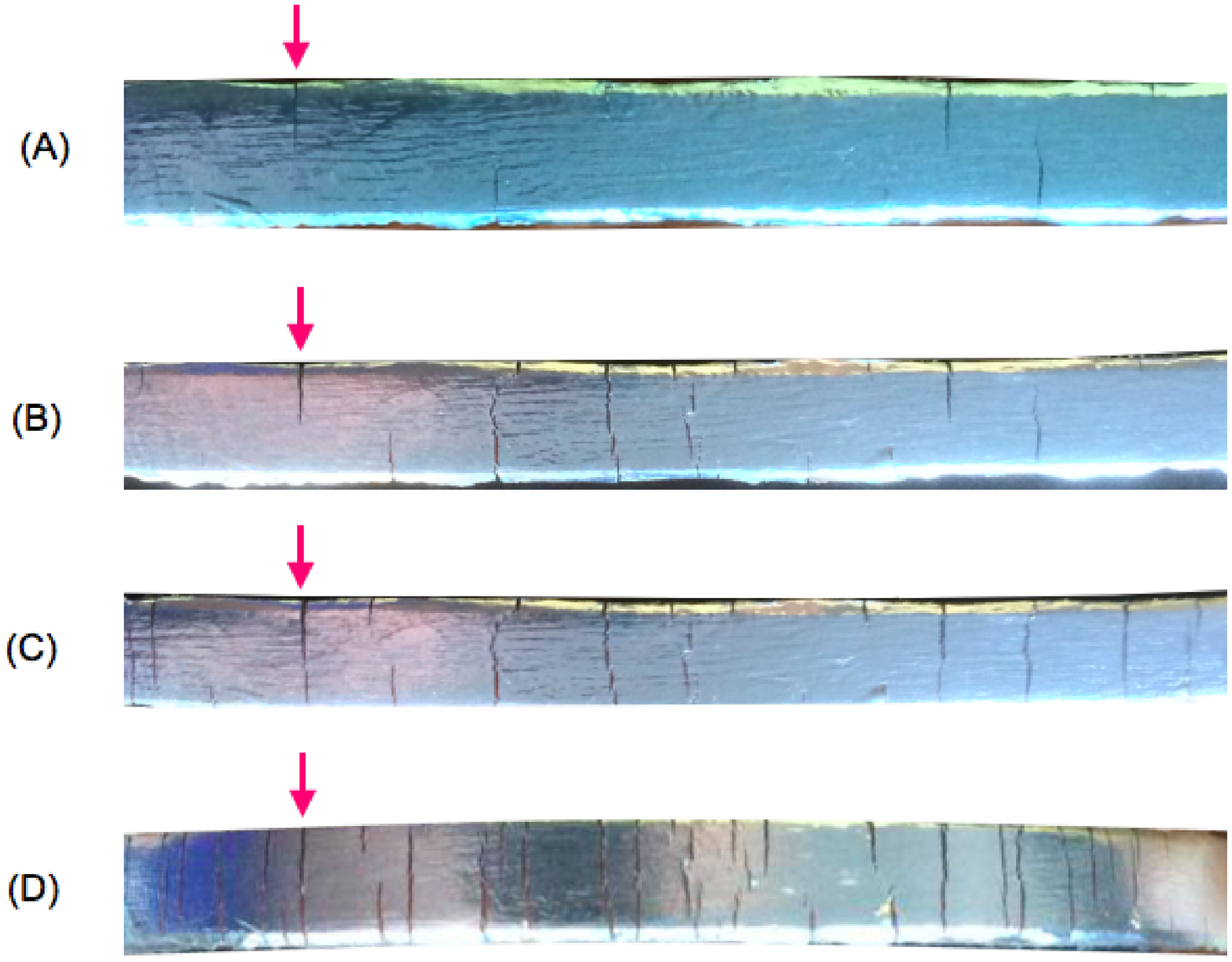}}\centerline{\includegraphics[scale=0.46]{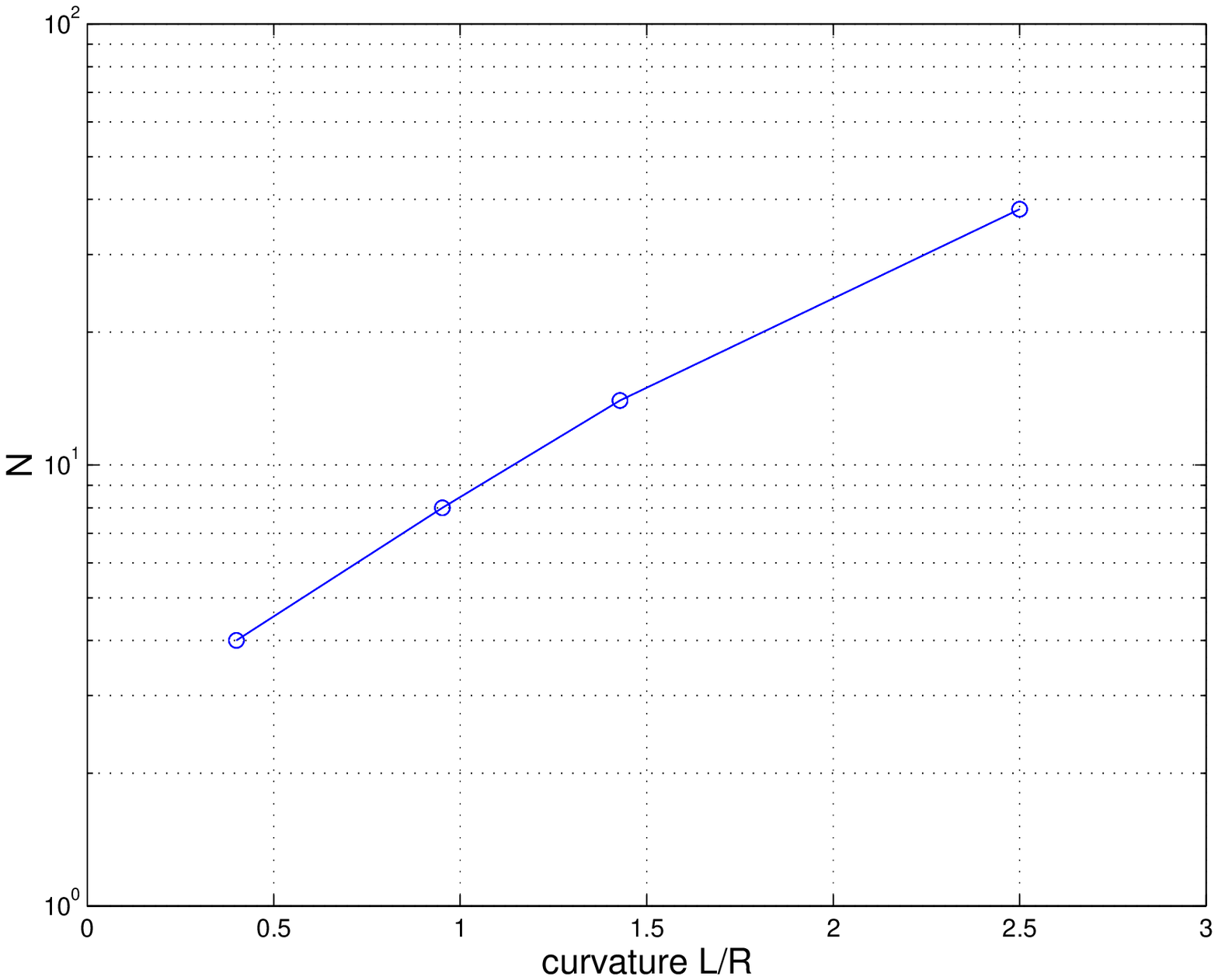}}
\caption{The formation of shear induced cracks in aluminum foil glued to a flexible rod of length $L$ subject to bending,
increasing from A-D ({\em top}). For reference, arrows point to the same crack at different degrees of bending.
The number of cracks increases essentially exponentially with curvature ({\em bottom}), representing a driving
deformation on large scales.}
\label{fig:GC3}
\end{figure}
\begin{figure}
\centerline{\includegraphics[scale=0.44]{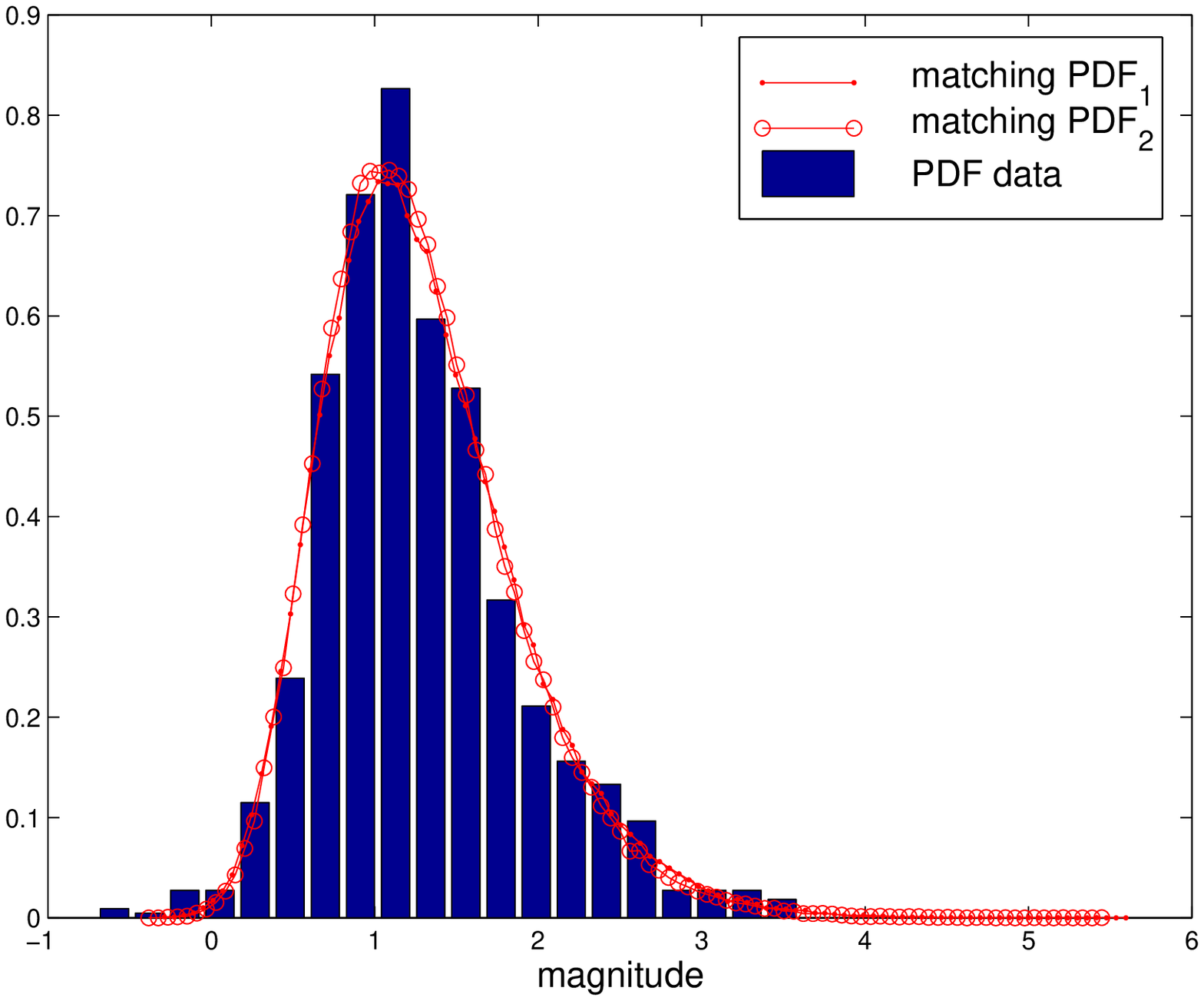}}
\centerline{\includegraphics[scale=0.44]{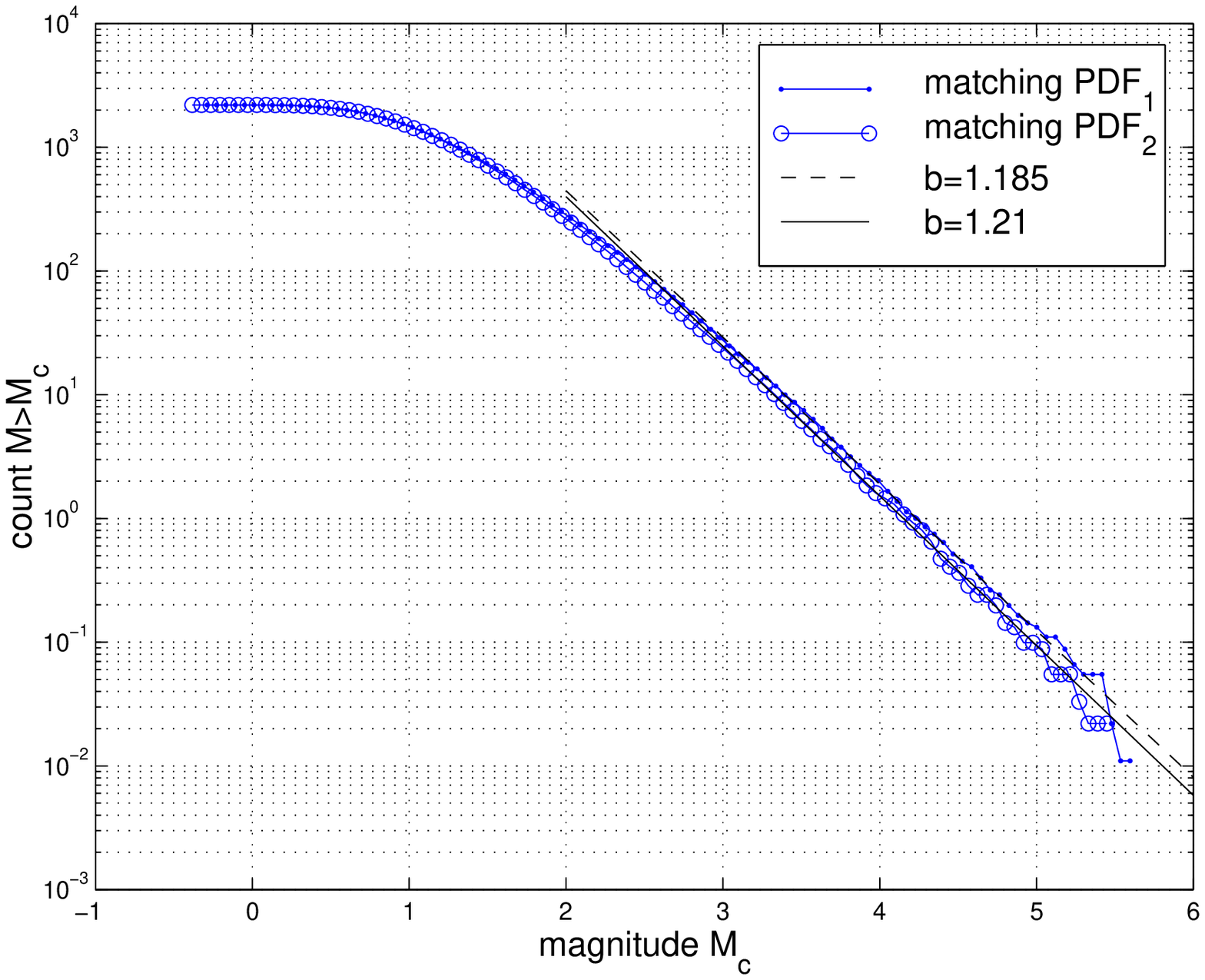}}
\caption{
({\em Top.}) The mean and standard deviation of all events in 2001-2014 are $(1.2876,0.6131)$ and, respectively, $(1.2706,0.5870)$,
where the event over the latter period show no discernible evolution in time. The observed distribution is matched by the distribution of 
maxima taken from large-$n$ samples taken from a Gaussian distribution. It shows an exponential tail at large magnitude, here for two 
matches PDF$_{i}$ reproducing the standard deviation of all events ($i=1$) and those since 2001 $(i=2)$. 
({\em Bottom.}) Expected cumulative event counts as a function of magnitude through 2025 based on
an anticipated additional 2200 events. Included are the tangents describing the exponential tail with negative slopes 
$b\simeq 1.2.$ A similar analysis on data up to 2013 shows $b=1.057$ and $b=1.138$ for PDF$_1$ and PDF$_2$, respectively,
indicative of an uncertainty of about 10\% in $b$.}
\label{fig:GC4}
\end{figure}

\section{Parameter-free matching of the magnitude distribution}

We next consider an earthquake event to be the sudden release of energy in a cluster of essentially contemporaneous events. In this case, the magnitude of the event is naturally attributed to the largest $\epsilon_c$ (with the largest $\eta$) in the cluster (cf. the composite source model of \citep{fra91}). 

The distribution of maxima in large $n$ samples of an underlying normal distribution of $\epsilon_c$ has an intrinsic positive 
skewness, e.g., by a Monte Carlo simulation (cf. Fig. 3 in \citep{van14}), which accurately matches in the observed distribution of the Groningen earthquake events in Fig. 1. The distribution of these maxima can be matched the observed distribution, here by matching their standard deviations. Fig. 4 shows two matching distributions of maxima of clusters of $n=2000$ tries taken from normal distributions of $\epsilon_c$. The matching distributions have exponential decay in the limit of large magnitude with a canonical Gutenberg-Richter $b$ value around 1 \citep{ish39,gut44,gut54,sch05}. Our mechanism obtains a value of $b=1.1-1.2$ shown in Fig. 7.

Our matching probability density functions (PDF) reproduce the mean and standard deviation of the PDF of the survey of observed magnitudes, not to a selected tail of magnitudes beyond a choice of cut-off. This procedure removes the cut-off as a free parameter, and ensures stability in estimating the tail at large magnitudes, which would otherwise be notoriously uncertain due to small number statistics. 

\section{Summary and outlook}

We report on a bimodal distribution of events in the US since 2013, where the new peak around magnitude M1 is very similar to the distribution of induced events in the Groningen field. Most likely, the new peak in the US is of anthropogenic origin.

The induced events in The Netherlands reveal an exponential growth rate with a doubling time of 6.24 years
and $b\simeq1.1-1.2$. In light of the observed formation of a shallow bowl due to a persistent rate of subsidence,
we explain this result by an exponential relation in the number of cracks or slides as a function of dimensionless curvature,
setting the leading order shape of the as a moving boundary in the elliptic problem for the internal strains in and around
the reservoir. A slow large scale creep, delayed relative to production, is consistent with a striking lack of a correlation of 
yearly subsidence with gas production and reservoir pressure. The origin of the slow creep, however, remains mysterious.

The PDF of earthquake magnitudes has a positive skewness which is well approximated by the distribution of maxima in clusters of events, taken as large $n$ samples from a normal distribution. In taking into account the full survey data, this estimate produces a $b$ value with a relatively moderate variation of about 10\%, somewhat tighter than obtainable by Gutenberg-Richter fits to low event counts in tails beyond a magnitude cut-off (e.g \citep{dos13}). The exponential tail in the limit of large magnitude combined with the anticipation additional 2200 events up to 2025 points to events likely up to magnitude 4.2 but unlikely reaching magnitude 5.

The accurate fit of our model PDF to the data is perhaps surprising in light of the completeness problem in earthquake
surveys, whereby the observed event counts represent a lower bound on the true event counts. Equivalently, the 
effective surface area of large $M$ events is larger than that of small $M$ events. However, events in the Groningen area 
are limited to a region limited within $30\times30$ km$^2$. It appears that the Groningen survey is rather complete
for events around M1. The new peak B about M1 in the distribution of events in the US is a positive detection based on
an overwhelmingly large number of events, where only the details of the event counts may be affected by finite 
completeness in the ANSSI Comprehensive catalogue.

Our analysis shows a mixed message: the number of earthquakes is rapidly increasing with the potential for significant
continuing exposure to gradual degrading of the integrity of buildings and homes, even as the magnitude of earthquakes 
probably remains below 5. The regional impact is primarily economical and political and increasingly endemic as a 
psychological stressor on the local population, even as immediate risk to injury or death remains small. 
The lack of any correlation between gas exploitation and event rate points to a significant delay time, perhaps well 
over a decade, which challenges the formulation of effective policies and strategies to ameliorate this outlook in the 
immediate future.

The exponential growth highlighted in Groningen may be typical, since there appear to be similar epochs also in the 
large scale counts of seismic activity in the United States. Particularly striking are the events in group $B$ emerging
suddenly in 2013. Although its exceptional growth remains to be determined in the coming years, its distribution already
shows remarkable similarity to the Groningen field with a similar outlook, in regions where it overlaps with populated
areas (cf. \citep{eck06,pet08}).

We propose that an accurate hazard analysis of exposure of populated areas to induced earthquakes takes into 
account the exponential growth in event counts and the relatively high frequency spectrum of induced motions,
as they fall in the range of eigenfrequencies of around 0.1-25 Hz of homes and buildings. For instance, the Fourier 
spectrum of induced events at the Hellisheidi Geothermal Powerplang in Iceland \citep{hal12} peaks at 10 Hz. This
frequency is high compared to earthquakes of tectonic origin. It can be attributed to the relatively larger magnitude 
of the latter (Fig. 1) according to the scaling $f\propto M^{-1/3}$ of frequency with magnitude 
\citep{aki67,bun70,kam75,and03,aki80}. Frequent small magnitude events can hereby 
be disproportionally hazardous to homes and buildings over time.

\section{Acknowledgments.} The first author thanks Tae Woong Chung for stimulating discussions.

\section{Data and Resources}

USGS data in Fig. 1 are obtained from the ANSS Comprehensive Catalogue at 
http://earthquake.usgs.gov/earthquakes/search/ (last accessed October 15 2014). ANSS Data 
were downloaded on October 15 2014 using the following search parameters in the USGS online portal: 
{\em Magnitude}: (minimum,maximum)=(-1,10), {\em Depth}: (minimum,maximum)=(-1,4) km, 
{\em Geographic Region}: (South,North)=(25,50), (West,East)=(-125,-65). Multiple time intervals were used to cover 
1974-2014 because of the total number of 48747 events. The Supplementary material provides a complete list of ANSS data
(ANSS$_-$Table.pdf), a video clip of the magnitude histogram as a function of time (US500.mpg4) and the USGS 
search parameters (USGS$_-$input.pdf).

KMNI data in Fig. 1 are obtained from http://www.knmi.nl/seismologie/aardbevingen-nederland.html (last accessed
February 13 2014). Data were extracted from http://www.knmi.nl/seismologie/geinduceerde-bevingen-nl on February 13 2014.

NAM data in Fig. 2 are obtained from NAM Production profielen (2012-2036) at NL Olie en Gasportaal,
http://www.nlog.nl/nl/home/NLOGPortal.html (last accessed May 13 2014) and extracted from
www.nlog.nl/oil/oilGas/grafiek$_-$productie$_-$profielen$_-$2013-2037$_-$PUBLIC.xls on February 13 2014.




 
\end{document}